\documentclass[12pt]{article}
\usepackage{a4wide}
\usepackage{cite}
\usepackage{latexsym}
\usepackage{amsfonts}
\usepackage{amssymb,amsbsy}
\usepackage{bbm}
\usepackage{slashed}
\usepackage{subeqn}
\usepackage{nonfloat}
\usepackage{morefloats}
\usepackage[dvips]{graphicx}
\renewcommand{\theequation}{\thesection.\arabic{equation}}
\def\vec#1{\boldsymbol{#1}}

\def\ppLL{\ensuremath{\bar{\mathrm{p}}\mathrm{p}\to\overline{\Lambda}\Lambda}}

\makeatletter
\newlength{\earraycolsep}
\def\eqnarray{\stepcounter{equation}\let\@currentlabel%
\theequation
\global\@eqnswtrue\m@th
\global\@eqcnt\z@\tabskip\@centering\let\\\@eqncr
$$\halign to\displaywidth\bgroup\@eqnsel\hskip\@centering
$\displaystyle\tabskip\z@{##}$&\global\@eqcnt\@ne
\hskip 2\earraycolsep \hfil$\displaystyle{##}$\hfil
&\global\@eqcnt\tw@ \hskip 2\earraycolsep
$\displaystyle\tabskip\z@{##}$\hfil
\tabskip\@centering&\llap{##}\tabskip\z@\cr}
\makeatother
\catcode`\@=11
\def\eqalign#1{\null\,\vcenter{\openup\jot\m@th
\ialign{\strut\hfil$\displaystyle{##}$&$\displaystyle{{}##}$\hfil
     \crcr#1\crcr}}\,}
\catcode`\@=12
\catcode`\@=11
\def\Eqalign#1{\null\,\vcenter{\openup\jot\m@th
\ialign{\strut\hfil$\displaystyle{##}$&$\displaystyle{{}##}$\hfil
&&\qquad\strut\hfil$\displaystyle{##}$&$\displaystyle{{}##}$\hfil
     \crcr#1\crcr}}\,}
\catcode`\@=12
\def\S{\Sigma}

\def\acc{\check{\rho}}

\def\Ccal{{\cal C}}
\def\Dcal{{\cal D}}
\def\Scal{{\cal S}}
\def\Rpt{{\tilde R}}
\def\R{R}
\def\Rin{\eta}
\def\acc{\check\rho}
\def\une{\mathbbm{1}}

\def\R{R}
\def\D{{\cal D}}
\def\<{\langle}
\def\>{\rangle}
\let\tr=\Tr
\def\det{{\rm det}}
\def\moins{-{1\over2}}
\def\plus{+{1\over2}}

\def\acc{\check{\rho}}

\def\sv{{\vec{s}}}

\def\Pv{{\vec{P}}}
\def\Sv{{\vec{S}}}
\def\Jv{{\vec{J}}}

\def\pv{{\vec{p}}}

\def\kv{{\vec{k}}}
\def\nv{{\vec{n}}}

\def\Pv{{\vec P}}

\def\xu{\hat{\vec{x}}}
\def\yu{\hat{\vec{y}}}
\def\zu{\hat{\vec{z}}}
\def\acc{\check{\rho}}

\def\plus{|+\rangle}
\def\moins{|-\rangle}      
\def\Xrond{{\rm X}\raise1.8ex\hbox{\kern-0.6em\hbox{$\circ$}}\ }
\def\krond{{\bf k}\raise1.8ex\hbox{\kern-0.6em\hbox{$\circ$}}\ }
\def\Mcal{{\mathcal{M}}}

\def\Ocal{{\mathcal{O}}}
\def\be{\begin{equation}}

\def\ee{\end{equation}}

\def\barr{\begin{eqnarray}}
\def\earr{\end{eqnarray}}

\def\sq{\raise -2pt\hbox{\Large $\Box$}}

\def\disp{\raise -1pt\hbox{\large$\otimes$}}
\def\trv{ \raise -2pt \hbox{\large$\nabla$}}
\def\trh{\raise -2pt \hbox{\rotatebox{90}{\hbox{\large$\Delta$}}}}

\begin{document}
\begin{center}
{\bfseries 
CLASSICAL AND QUANTUM CONSTRAINTS IN SPIN PHYSICS}
\vskip 5mm 
{X. Artru}
\vskip 5mm
{\small 
{\it Institut de Physique Nucl\'eaire de Lyon, Universit\'e de Lyon,}\\
{\it CNRS- IN2P3 and Universit\'e Lyon 1, 69622 Villeurbanne, France}\\
{\it E-mail: x.artru@ipnl.in2p3.fr}
}
\end{center}
\vskip 8mm
\begin{abstract}
Constraints on spin observables coming from discrete symmetries such as P, C, T and identical particles may be divided in two types: 1) classical ones, which insure the invariance of the cross sections under the symmetry operation; 2) non-classical ones, which can only be obtained at the level of amplitudes. 
Similarly, positivity constraints can be divided into classical and non-classical constraints. The former insure the positivity of the cross section for arbitrary individual polarisations of the external particles, the latter extend this requirement to the case of entangled external spins. 
The domain of classical positivity is shown to be dual to the domain of separability. 
\end{abstract}

\vskip 8mm
\setcounter{equation}{0}

\section{The spin observables} 

We consider the polarised $2\times2$ reaction
\be\label{reaction}
A+B \to C+D\,,
\ee
where $A$, $B$, $C$ and $D$ are spin one-half particles. Let us recall some of the formalism presented in 
\cite{domaines,larevue}. The fully polarised differential cross section of (\ref{reaction}) reads
\be\label{basic:eq:dcs}
{d\sigma \over d\Omega} = I_0 \ F\left( \Sv_A,\Sv_B,\check \Sv_C,\check \Sv_D \right)\,,
\ee
where $F$ contains the spin dependence. $\Sv_A$ and $\Sv_B$ are the polarisation vectors of the initial particles ($|\Sv| \le 1$). $\check \Sv_C$ and $\check \Sv_D$ are pure polarisations ($|\check\Sv|=1$)  \emph{accepted} by an ideal spin-filtering detector. They must be distinguished from the \emph{emitted} polarisations $\Sv_C$ and $\Sv_D$ of the final particles. These ones depend on the polarisations of the incoming particles, e.g.,
\be\label{outgoingPol}
\Sv_C = \nabla_{\check\Sv_C} F(\Sv_A,\Sv_B,\check\Sv_C,\check\Sv_D=0) \,/\,F(\Sv_A,\Sv_B,\check\Sv_C=0,\check\Sv_D=0) 
\ee
$F$ is given in terms of the \emph{Cartesian reaction parameters} \cite{Bourrely:1980mr} by 
\be
F\left( \Sv_A,\Sv_B,\check \Sv_C,\check \Sv_D \right)
= C_{\lambda\mu\nu\tau} \ S^\lambda_A \, S^\mu_B \, \check S^\nu_C \, \check S^\tau_D
\,.
\label{FCart}
\ee
In the right-hand side the $\Sv\,$'s are promoted to four-vectors with $S^{0} = 1$.
The indices $\lambda, \mu, \nu, \tau$, run from 0 to 3,
whereas latin indices $i$, $j$, $k$, $l$, take the values 1, 2, 3, or $x$, $y$, $z$.
A summation is understood over each repeated index.
$S^x, S^y, S^z$ are measured in a triad of unit vectors
$ \{ \xu, \yu, \zu \} $ which may differ from one particle to the other. A standard choice is to take $\zu$ along the particle momentum and $\yu$ common to all particles and normal to the scattering plane. 
Conversely we have
\be\label{basic:eq:corr-param}
C_{\lambda\mu\nu\tau} = \tr\{\,\mathcal{M}\,\left[\sigma_\lambda(A)\otimes\sigma_\mu(B)\right]\,\mathcal{M}^\dagger
\left[\sigma_\nu(C)\otimes\sigma_\tau(D)\right]\,\}\ /\ \tr\{\,\mathcal{M}\,\mathcal{M}^\dagger\,\}\,,
\ee
which will be symbolically abbreviated as a sort of expectation value:
\be
(\lambda\mu|\nu\tau)\equiv C_{\lambda\mu\nu\tau}=\langle\sigma_\lambda(A)\,\sigma_\mu(B)\,\sigma_\nu(C)\,\sigma_\tau(D)\rangle~,\label{basic:eq:corr-param-bis}
\ee
with $\sigma_0=\mathbbm{1}\equiv\pmatrix{1&0\cr0&1}$. 

\medskip
\noindent
{\bf 3. Classical and quantum constraints for parity}

\medskip
The scattering plane is a symmetry plane for the reaction (\ref{reaction}), which is therefore symmetric under under the {mirror reflection} 
\begin{equation}
	\Pi = P \,\exp(-i\pi \Jv_y)~.
\end{equation}
If parity is conserved the matrix amplitude $\Mcal$ of  $A+B\to C+D$ fulfils:
\be
	\Mcal = \left(\Pi_C\otimes\Pi_D\right)^{-1}\Mcal\left(\Pi_A\otimes\Pi_B\right)~.
\label{basic:eq:P-ampli}
\ee
For one fermion, $\Pi=-i\eta\,\sigma_y$, where $\eta$ is the intrinsic parity of the fermion.
Applying this equation to both $\Mcal$ and $\Mcal^\dagger$ in (\ref{basic:eq:corr-param}) one obtains the \emph{classical parity rule} 
\be\label{basic:eq:P-class-O}
\left\langle\sigma_\lambda(A)\,\sigma_\mu(B)\,\sigma_\nu(C)\sigma_\tau(D)\,\right\rangle
=
\left\langle\sigma_\lambda^\Pi(A)\,\sigma_\mu^\Pi(B)\,\sigma_\nu^\Pi(C)\,\sigma_\tau^\Pi(D)\right\rangle
~,
\ee
where $\Ocal^{\Pi}$ denotes the reflected operator $\Pi\ \Ocal\ \Pi^{-1}$. For the Pauli matrices, the reflection reads
\be
	(\sigma_0,\,\sigma_x,\,\sigma_y,\,\sigma_z) \to (\sigma_0,\,-\sigma_x,\,\sigma_u,\,\sigma_z)~.
	\label{basic:eq:sigma-parity}
\ee
The multi-spin observable $O_{\lambda\mu\nu\tau}=O_\lambda(A)\otimes O_\mu(B)\otimes O_\nu(C)\otimes O_\tau(D)$ is
$\Pi$-odd if it contains an odd number of $\Pi$-odd Pauli matrices, otherwise it is $\Pi$-even.
The ``classical'' rule reads:  

\medskip\centerline{ \emph{If parity is conserved, all $\,\Pi$-odd observables vanish.}} 

\medskip
\noindent
For instance, $(z0|y0)=0$, but $(00|y0)\ne0$. 
This rule roughly reduces by a factor 2 the number of observables. It does not depend on the intrinsic parity of the particles. It just expresses a classical requirement of reflection symmetry at the level of polarised \emph{cross sections}. 

Applying (\ref{basic:eq:P-ampli}) only to $\Mcal$ or to $\Mcal^\dagger$ in (\ref{basic:eq:corr-param}) one  obtains the \emph{non-classical} parity constraint
\be
\left\langle\sigma^\lambda_A\,\sigma^\mu_B\,\sigma^\nu_C\,\sigma^\tau_D\right\rangle
=
\left\langle
\left(\Pi_A\,\sigma^\lambda_A\right)\ \left(\Pi_B\,\sigma^\mu_B\right)\ \left(\sigma^\nu_C\,\Pi_C^{-1}\right)\ \left(\sigma^\tau_D\,\Pi_D^{-1}\right)\right\rangle
~,\label{basic:eq:P-quant-O}
\ee
with $\Pi=-i\eta\,\sigma_y$. For the ${1\over2}^+$ baryons one can choose $\eta=i$ so that 
$\Pi=\Pi^{-1}=\sigma_y$ and 
\be\label{basic:eq:P-quant-O-bis}
\Pi\ \left(\,\sigma_0,\ \sigma_x,\ \sigma_y,\ \sigma_z\,\right)
=\left(\sigma_y,\ -i\sigma_z,\ \sigma_0,\ i\,\sigma_x\,\right)~.
\ee
For a pseudoscalar meson, $\Pi=-1$. For example in $\pi+N\to K+\Lambda$ on gets
\be
(y|y)=(0|0)\,,\quad (0|y)=(y|0)\,.
\ee
Clearly the first of these constraints, which relates a polarised cross section to an unpolarised one, cannot be obtained by classical parity arguments. 
The non-classical parity constraints in the case of spin one-half particles are known as the \emph{Bohr identities} \cite{Bohr:1959}. 
Non-classical parity rules depend on the intrinsic parities. They yield linear identities between the $\Pi$-even observables and reduce the number of independent correlation parameters roughly by another factor 2. For instance, in $\pi^0$ decay, the classical parity rule tells that the linear polarisations of the two gamma's are either parallel or orthogonal (not, e.g. at $\pi/4$). The analogue of (\ref{basic:eq:P-quant-O}) for photons selects the orthogonal solution. 

The subdivision in constraints of the (\ref{basic:eq:P-class-O}) and (\ref{basic:eq:P-quant-O}) types, both for parity and time-reversal, has already been made in literature (see Appendix 3.D. of \cite{Bourrely:1980mr}). Here we  point out the ``classical'' versus ``non-classical'' or ``quantum'' characters of these two types. Inclusive reactions have only ``classical'' parity constraints, since the intrinsic parity of the undetected particles can take both signatures. 

Similar divisions in classical versus non-classical constraints can be made for other symmetries like charge conjugation, time reversal and permutation of identical particles.

\bigskip
\noindent
{\bf 4. Classical positivity constraints}

\medskip
The cross section (\ref{basic:eq:dcs}) has to be positive for arbitrary \emph{independent} polarisations of the external particles, that is to say
\be\label{class-posit}
F\left( \Sv_A,\Sv_B,\check \Sv_C,\check \Sv_D \right)\le 1
\quad{\rm for}\quad 
\Sv_A,\ \Sv_B, \ \check\Sv_C,\ \check\Sv_D \in \hbox{unit ball}\ |\Sv|\le1\,.
\ee
%
An equivalent condition is that the polarisation of, for instance, outgoing particle $C$ for given $\Sv_A$, $\Sv_B$, and imposed $\check\Sv_D$, 
\be\label{outgoingPol'}
\Sv_C(\Sv_A,\Sv_B,\check\Sv_D)  = \nabla_{\check\Sv_C} F(\Sv_A,\Sv_B,\check\Sv_C,\check\Sv_D) \,/\,F(\Sv_A,\Sv_B,\check\Sv_C=0,\check\Sv_D) 
\ee
lies in the unit ball $|\Sv_C|\le1$ for any $\Sv_A$, $\Sv_B$ and $\check\Sv_D$. For instance  in $\pi+N\to K+\Lambda$ the inequalities
\be
(C_{0x}\pm C_{zx})^2 +(C_{0y}\pm C_{zy})^2+(C_{0z}\pm C_{zz})^2\le (C_{00}\pm C_{z0})^2
\ee
insure that the $\Lambda$ polarisation does not exceed 1 when the nucleon polarisation is longitudinal. 

The condition (\ref{class-posit}) defines a convex \emph{classical positivity domain} $\Ccal$ in the space of the Cartesian reaction parameters. As we shall see, it is a necessary but \emph{not sufficient} positivity condition. 

\bigskip
\noindent
{\bf 5. Quantum positivity constraints}

\medskip
All spin observables of reaction (\ref{reaction}) can be encoded in the \emph{cross section matrix} $R$, or its partial transpose $\Rpt$, defined by
\be\label{basic:eq:CSdM}\eqalign{  
\langle c,d | \mathcal{M}| a,b \rangle\,\langle a',b' | \mathcal{M}^\dagger| c',d' \rangle 
&=\langle a',b';c',d'|\R|a,b\,;c,d\rangle\cr
&=\langle a',b';c,d |\Rpt|a,b\,;c',d'\rangle
\,.}\ee
The transposition linking $\Rpt$ to $R$ bears on the final particles. 
The diagonal elements of $R$ or $\Rpt$ are the fully polarised cross sections when the particles are in the basic spin states. By construction, $\R$ (but not necessarily $\Rpt$) is \emph{semi-positive definite}, that is to say $\langle\Psi|\R|\Psi\rangle\ge0$ for any $\Psi$. 

Equations (\ref{basic:eq:dcs}), (\ref{FCart}) and (\ref{basic:eq:corr-param}) can be rewritten as:
%
\be\label{basic:eq:dpcsR}\eqalign{%
{d\sigma\over
d\Omega}\left(\rho_A,\rho_B,\acc_C,\acc_D \right) &=
\tr\{\Rpt\ [\rho_A\otimes\rho_B\otimes\acc_C\otimes\acc_D]\,\}~,\cr
\phantom{{1\over1}} C_{\lambda\mu\nu\tau} &= \tr\{\Rpt\left[
\sigma_\lambda(A)\otimes\sigma_\mu(B)\otimes\sigma_\nu(C)\otimes\sigma_\tau(D)\right]\}\, /\,\tr \Rpt~,}
\ee 
with $\rho={1\over2}(\une+\Sv\cdot\vec\sigma)$, $\ \acc={1\over2}(\une+\check\Sv\cdot\vec\sigma)$. 
The last equation of (\ref{basic:eq:dpcsR}) can be inverted as 
\be\label{basic:eq:csdm}\eqalign{%
{\Rpt}_1\equiv(2^4/\tr\Rpt)\ \, \Rpt\ &=\ C_{\lambda\mu\nu\tau}\ \sigma_\lambda(A)\otimes\sigma_\mu(B)\otimes\sigma_\nu(C)\otimes\sigma_\tau(D)~,\cr
{\rm or}\quad {\R}_1\equiv
(2^4/(\tr\R)\ \, \R\ &=\ C_{\lambda\mu\nu\tau}\ \sigma_\lambda(A)\otimes\sigma_\mu(B)\otimes\sigma^t_\nu(C)\otimes\sigma^t_\tau(D)~.}
\ee
The matrix ${\Rpt}_1$ is normalised to have the same trace as the unit matrix and is directly obtained from $F$ replacing the $S^\mu$'s by $\sigma^\mu$'s. It allows to calculate the cross section for \emph{entangled} initial states, replacing $\rho_A\otimes\rho_B$ by $\rho_{A+B}$ in (\ref{basic:eq:dpcsR}), as well as the joint density matrix of $C$ and $D$:
\be
\rho_{C+D}=\tr_{A,B}\{\,\Rpt\ [\rho_A\otimes\rho_B]\,\}
/\tr\{\Rpt\,[\rho_A\otimes\rho_B]\,\}~.
\ee
The single polarisation of particle $C$ can then be obtained by $\rho_{C}=\tr_{D}\{\rho_{C+D}\}$, 
in place of (\ref{outgoingPol}). 

The semi-positivity of $\R$ leads to \emph{quantum positivity} constraints on the Cartesian reaction parameters which are stronger than the classical ones. 
Suppose, for instance, that 
\be
F(\Sv_A,\Sv_B,0,0)\propto 1+c\,\Sv_A\cdot\Sv_B\,.
\ee
Then $\R\propto\mathbbm{1}+c\,\vec\sigma_A\cdot\vec\sigma_B$, where $\vec\sigma_A\cdot\vec\sigma_B\equiv\sum_{i=1}^3\sigma_A^i\otimes\sigma_B^i$, and the initially polarised cross section is 
\be\label{basic:eq:dpcsR1}
{d\sigma\over d\Omega}\left(\rho_{A+B}\right) = 
\tr\{\Rpt\ \left[\rho_{A+B}\otimes\une_{C+D}\right]\,\}
\propto\tr\{(\une+c\,\vec\sigma_A\cdot\vec\sigma_B)\,\rho_{A+B}\,\}~.
\ee
For uncorrelated $\Sv_A$ and $\Sv_B$ one has $d\sigma/d\Omega\propto1+c\,\Sv_A\cdot\Sv_B\ge0$, therefore classical positivity is fulfilled for $c\in[-1,+1]$. However, if $A$ and $B$ form a singlet spin state, of density matrix $\rho_{A+B}={1\over4}\,(\une-\vec\sigma_A\cdot\vec\sigma_B)$, then $d\sigma/d\Omega$ is positive only for $c\in[-1,+1/3]$.

The occurrence of a negative cross section comes from the non-positivity of $\mathbbm{1}+c\,\vec\sigma_A\cdot\vec\sigma_B$ for $c>1/3$. This non-positivity was revealed by an entangled initial state (the spin singlet state). This example shows that positivity has to be tested not only with factorised (or separable states), but also with entangled ones. 

Similarly, a final spin correlation of the form $F(0,0,\check\Sv_C,\check\Sv_D)=1+c\,\check\Sv_C\cdot\check\Sv_D$ is classically allowed for $c\in[-1,+1]$, but quantum-mechanically for $c\in[-1,+1/3]$ only. As a check rule, ``quantum mechanics does not allow fully parallel spins''. 
These examples have a crossed symmetric counterpart: a spin transmission between $A$ and $C$ of the form
\be
F(\Sv_A,0,\check\Sv_C,0)=1+c\,\Sv_A\cdot\check\Sv_C\,
\ee
is classically allowed for $c\in[-1,+1]$, but quantum-mechanically for $c\in[-1/3,+1]$ only. For $c<-1/3$ the cross section matrix is non-positive and this can be revealed by an ``entangled state in the $t$-channel''. The corresponding check rule is ``quantum mechanics does not allow full spin reversal''. 
The lesson of these examples is that positivity  has to be tested with classical and \emph{entangled} states in the \emph{direct} and \emph{crossed} channels. 

An example of non-classical positivity constraint is the the Soffer inequality~\cite{Soffer}:
$$
2 \delta q(x) \le q(x) + \Delta q(x) \,
$$
between the quark helicity- and transversity distributions  $\Delta q(x)$ and $\delta q(x)$. 

\bigskip
\noindent
{\bf 6. Domains of quantum positivity, classical positivity and 
separability}

\medskip
As we have seen one can distinguish a \emph{classical positivity domain} which is larger than the true or 
\emph{quantum} positivity domain. To have a more precise idea about the differences between these two domains, let us study the constraints on the initial spin observables only. For this purpose we introduce the matrix 
\be\label{basic:eq:reducedAC} 
\Rin_{A+B} = \tr_{C,D}\left[\R/(\tr\R)\right]~
\ee
obtained by taking the partial trace over the final particles and renormalising to unit trace. Like $\R$, $\Rin_{A+B}$ has to be (semi-)positive. The initially polarised cross section reads
\be\label{basic:eq:dpcsR'}
{d\sigma\over d\Omega}\left(\rho_{A+B}\right) =
\tr\{\Rin_{A+B}\ \rho_{A+B}\}
~,
\ee
Classical positivity requires $\tr 
\{\Rin_{A+B}\ (\rho_A\otimes\rho_B)\}\ge0$ for any individual density matrices $\rho_A$ and $\rho_B$.  More generally
\be\label{furth:eq:eq:classic+}
\tr \{\Rin_{A+B}\, \rho_{A+B}\}\ge0 \quad \hbox{for any 
\emph{separable}} \ \rho_{A+B}\,,
\ee
whereas quantum positivity requires
\be\label{furth:eq:quant+}
\tr \{\Rin_{A+B}\,\rho_{A+B}\}\ge0 \quad \hbox{for any separable 
\emph{or entangled}} \ \rho_{A+B}\,.
\ee
One can say that the classical positivity domain $\mathcal{C}$ is 
\emph{dual} to the separability domain $\mathcal{S}$ in the sense 
that $\tr \{\Rin\,\rho\}\ge0$ for any pair 
$\{\Rin\in\mathcal{C}\,,\rho\in\mathcal{S}\}$. 
As for the quantum 
positivity domain $\mathcal{D}$, it is dual to itself. We have
\be\label{furth:eq:SDC}
\mathcal{S}\subset\,\mathcal{D}\,\subset\mathcal{C}\,,
\ee 
these three domains being convex. 

\begin{figure}[!thc]
\begin{minipage}{.59\textwidth}
\centerline{\includegraphics[width=.95\textwidth]{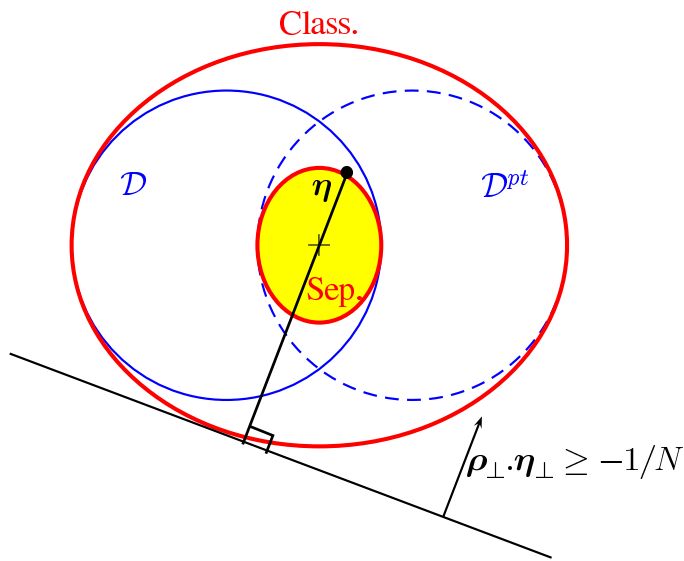}} 
\end{minipage}
\
\begin{minipage}{.39\textwidth}
\caption{\label{furth:fig:YYY}%
Schematic shapes of the classical positivity domain in the $\rho_\perp$ space. 
($\mathcal{C}\equiv\mathcal{C} lass.$), the separability domain 
($\mathcal{S}\equiv\mathcal{S} ep.$) and the true positivity domain 
$\mathcal{D}$.
The dashed contour indicates the domain $\mathcal{D}^{pt}$ where the partial transform is positive. 
A matrix $\Rin$ of the boundary $\partial\mathcal{S}$ is represented 
together with its reciprocal polar line $\vec\rho_\perp\cdot\vec\eta_\perp=1/N$, which is tangent to 
$\partial\mathcal{C}$. 
}
\end{minipage}
\end{figure}

Let us take the traceless part $\rho_\perp=\rho-\une/N$ of $\rho$, where $N=\tr(\une)$ is the dimension of the $A+B$ spin space, and introduce the Euclidian scalar product $\vec\Rin_\perp\cdot\vec\rho_\perp=\tr(\Rin_\perp\rho_\perp)$
where $\vec\rho_\perp$ is considered as a $N^2-1$ dimensional vector. The duality between $\mathcal{C}$ and $\mathcal{S}$ can be expressed as
\be\label{furth:eq:polar-recip} 
\vec{\rho}_\perp\cdot\vec\Rin_\perp\ge-1/N\,,\quad\rho\in\mathcal{C}\,,\ 
\Rin\in\mathcal{S}\,.
\ee
Figure \ref{furth:fig:YYY} schematises the properties (\ref{furth:eq:SDC}) and (\ref{furth:eq:polar-recip}) in the $\vec{\rho}_\perp$ space. Equation (\ref{furth:eq:polar-recip}) means that the boundaries $\partial\mathcal{C}$ and $\partial\mathcal{S}$ of the two domains are \emph{polar reciprocal} 
of each other: when $\vec\Rin_\perp$ moves on $\partial\mathcal{S}$, 
the reciprocal plane in $\vec\rho_\perp$ space defined by 
$\vec\rho_\perp.\vec\Rin_\perp=-1/N$ envelops $\partial\mathcal{C}$, 
as shown in Fig.~\ref{furth:fig:YYY}. 
Also shown in this figure is the symmetry between $\mathcal{D}$ and the domain $\mathcal{D}^{pt}$ where the partial transform $\rho_{A+B}^{pt}$ of $\rho_{A+B}$ is positive, the transposition concerning either $A$ or $B$. Indeed separability \cite{Peres,Horodecki} and classical positivity are preserved under partial transposition and we have
\be\label{furth:eq:SDCpt}
\mathcal{S}\subset\,\mathcal{D}^{pt}\,\subset\mathcal{C}\,.
\ee 
%
%

\begin{figure}[here]
\begin{minipage}{.59\textwidth}
\centerline{\includegraphics[width=.95\textwidth]{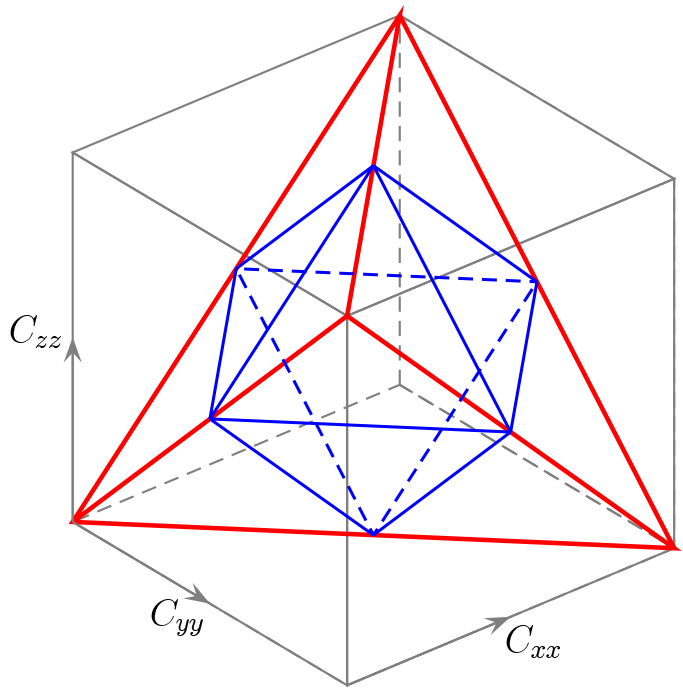}}
\end{minipage}
\
\begin{minipage}{.39\textwidth}
\caption{\label{furth:fig:ZZZ}%
Classical positivity domain (cube), true positivity domain 
(tetrahedron) and separability domain (octahedron) for the triple 
$\{C_{xx}\,,C_{yy}\,,C_{zz}\}$ of observables. }
\end{minipage}
\end{figure}

The duality between $\mathcal{C}$ and $\mathcal{S}$ may still be 
visible with a subset of observables. For instance, for a two-fermion 
system of density matrix 
$\rho_{A+B}={1\over4}\,C_{\mu\nu}\,\sigma_\mu(A)\otimes\sigma_\nu(B)$, 
the classical positivity domain of the triple 
$\{C_{xx}\,,C_{yy}\,,C_{zz}\}$ is the whole cube $[-1,+1]^3$, the 
quantum positivity domain is the tetrahedron defined by
\be\label{furth:eq:xxx}
C_{xx}-C_{yy}-C_{zz}\le1\quad\hbox{and circular permutations,}\quad 
C_{xx}+C_{yy}+C_{zz}\le1\,,
\ee
and the separability domain, an octahedron, is the intersection of 
the tetrahedron with its mirror figure. One can see on 
Fig.~\ref{furth:fig:ZZZ} the polar reciprocity (edge 
$\leftrightarrow$ edge) and (summit $\leftrightarrow$ face) between 
the cube and the octahedron. Related results are found in \cite{Bertlmann}. 

\bigskip
\noindent
{\bf 7. Outlook}
\medskip

We have qualified as \emph{classical} the symmetry and positivity constraints which can be derived by classical arguments concerning the polarised cross sections for separate polarisations of the external particles. 
Working at the level of amplitudes, or of the cross section matrix, one obtains quantum constraints which in many cases are stronger than the classical ones, therefore called non-classical. The number of non-classical constraints is expected to decrease when only part of the external particles are polarised or analysed, and in fact, there is no non-classical parity constraint for inclusive reactions. 
The weakening of non-classical constraints when part of the information is lost or discarded has some similarity with decoherence. Nevertheless some non-classical positivity constraints, for instance the Soffer inequality, still remain in the inclusive case.

A duality has been established between the domains of separability $\mathcal{S}$ and classical positivity $\mathcal{C}$ . 
In the space of the traceless components $\vec{\rho}_\perp$, the boundary $\partial\mathcal{S}$ and $\partial\mathcal{C}$ of these domains are polar reciprocal of each other. The boundary of $\mathcal{C}$ can be determined by algebraic equations using (\ref{class-posit}). This may offer a method for the long-standing problem of determining $\mathcal{S}$. 

\paragraph{Acknowledgements}
The author thanks M. Elchikh, O.V. Teryaev, J.M. Richard and J. Soffer for help, useful discussions and comments. 


%

\begin{thebibliography}{99}
%

\bibitem{domaines} X. Artru, J.M. Richard and J. Soffer, submitted to this Proceedings. 

\bibitem{larevue} X. Artru, M. Elchikh, J.M. Richard, J. Soffer and O. V. Teryaev, submitted to Phys. Reports.

\bibitem{Bourrely:1980mr} C. Bourrely, J. Soffer and E. Leader, \ Phys.\ Rept.\ {\bf 59} (1980) 95.

\bibitem{Bohr:1959} A. Bohr, \ Nucl.\ Phys.\ {\bf 10} (1959) 486.

\bibitem{Artru:2004jx} X. Artru and J.M. Richard, \ Phys.\ Part.\ Nucl.\ {\bf 35} (2004) S126, Proc. Dubna Spin Workshop.

\bibitem{Soffer}
J. Soffer, Phys. Rev. Lett. {\bf 74} (1995) 1292.

\bibitem{Peres} A. Peres, \ Phys.\ Rev.\ Lett.\ {\bf 77} (1996) 1413.

\bibitem{Horodecki} M. Horodecki, P. Horodecki and R. Horodecki, Phys. Lett. A {\bf 233} (1996) 1.

\bibitem{Bertlmann} R. A. Bertlmann, H. Narnhofer and W. Thirring, \ Phys.\ Rev. A {\bf 66} (2002) 032319.
%
%
\end{thebibliography}
\end{document}